\documentclass[letter]{aa}  

\usepackage{natbib}
\usepackage{graphicx}
\usepackage{subcaption}
\usepackage{float}
\usepackage{hyperref}
\usepackage{soul}
\usepackage{txfonts}
\begin{document}

   \title{A Tale of NGC\,3785: The Formation of an Ultra-Diffuse Galaxy at the end of the Longest Tidal Tail}

   \author{Chandan Watts
          \inst{1,2}
          \and
          Sudhanshu Barway\inst{1}
          \and
          Omkar Bait\inst{3}
          \and
          Yogesh Wadadekar\inst{4}
          }

   \institute{Indian Institute of Astrophysics,
              II Block, Koramangala, Bengaluru 560 034, India.\\
              \email{chandan@iiap.res.in} \and Pondicherry University, R.V. Nagar, Kalapet, 605014, Puducherry, India \and SKA Observatory, Jodrell Bank, Lower Withington, Macclesfield, SK11 9FT, UK \and National Centre for Radio Astrophysics, TIFR, Post Bag 3, Ganeshkhind, Pune 411 007, India
             }

   \date{}

 
  \abstract
   {}
    {We present the discovery of an extended and faint tail observed in the isolated environment associated with galaxy NGC\,3785. This study additionally provides observational evidence supporting the formation of ultra-diffuse galaxies at the end of the tail.}
   {We utilized the Gnuastro software to detect and analyse the low surface brightness structures in the optical $g$, and $r$ bands using data from the Dark Energy Camera Legacy Survey. We created a detection map to identify the faint tail, and measured its length using cubic spline fitting. Additionally, we found 84 star-forming clumps along the tail and performed photometric analysis on the tail portion after applying a significance threshold on the signal-to-noise ratio .}
   {We have measured the projected length of the tail, which is $\sim$ 390 kpc. We propose that this tail arises from the interaction of the NGC\,3785 with a gas-rich galaxy, which ends up as ultra-diffuse galaxy at the end of the tail.}
   {}


   \maketitle
%
\section{Introduction}
Galaxy mergers and interactions play an essential role in the evolution of galaxies. These mergers can be classified into subtypes depending on the mass ratio between the less massive and more massive galaxy \citep{bottrell2023illustristng}. For instance, minor mergers have mass ratios < 0.25 and are more frequent than major mergers. Whenever galaxies interact, they produce tidal features depending upon the nature of the merger and the fraction of gas involved in this process. \cite{1972ApJ...178..623T,1992ApJ...393..484B} showed in their simulations that tidal bridges and long tails are generated during major mergers, while minor mergers result in tidal streams where the less massive galaxies fall into the more massive ones \citep{2010AJ....140..962M}.

This fascinating phenomenon occurs when two galaxies collide with each other, resulting in gas and stars being extracted in the form of tidal tails and streams \citep{2024arXiv240809898W}. \cite{1998ApJ...494..183M,1999MNRAS.307..162S} have shown in their studies that the morphology and length of the tidal tails significantly depends on the dark matter halo potential. It has been observed that these tidal tails are blue in color, and most of the gas accumulates near the tip of the long tails, which results in the formation of tidal dwarf galaxies \citep{2004A&A...427..803D} in gas-rich mergers. \cite{2016ApJ...832...90W} conducted a study on 461 galaxies with tidal tails in the redshift range of $0.2 < z < 1$. They found that these tidal tail galaxies generally originate from star-forming disk galaxies. 

Here, we present the galaxy NGC\,3785, which possesses a long extended tail. NGC\,3785 has S0 type morphology \citep{1991rc3..book.....D}, with redshift $\sim$ 0.03 \citep{2017ApJS..233...25A} corresponding to a distance of $\sim$132 Mpc from us (Table \ref{table:intro}). This galaxy has an extended tail with both straight and curved sections. It has a companion galaxy, {\it SDSS J113939.59+262036.6}, with the same redshift. These two galaxies are at a projected distance of $\sim$ 103 kpc. It is unlikely that interactions between these two are responsible for the tidal tail due to their significant separation.
 
This extended tail is visible in the DECaLS (Dark Energy Camera Legacy Survey), SDSS(Sloan Digital Sky Survey) and Pan-STARRS(Panoramic Survey Telescope and Rapid Response System) images. However, it is most clearly seen in the DECaLS observation due to the higher sensitivity of that survey. We describe the DECaLS observations in the next section. 
 
In this paper, we present the discovery of an extended tail and study its properties such as length, and surface brightness. In Section 2, we describe the data we used. In Section 3, we discuss the detection of tail and star-forming clumps associated with the tail. In Section 4, we present the findings obtained from the data and their connection with the previous studies. Section 5  summarizes the results. 

Throughout this paper, we used the Flat-$\Lambda$CDM model with $H_0=70$~km/s/Mpc, $\Omega_{m} = 0.286$ and $\Omega_{vac} = 0.714$ which corresponds to a cosmological scale of 0.602 kpc/$\arcsec$ at the redshift of  NGC\,3785.

\begin{table}[h]
\caption{Properties of the galaxy NGC\,3785. These parameters are taken from NED.}
 \label{table:intro}
$$ 
\begin{array}{p{0.5\linewidth}l}
    \hline
    \noalign{\smallskip}
    RA[J2000] &   11h39m32.891s   \\
    Dec[J2000] &   +26d18m08.18s          \\
    Morphological Type        &  S0  \\
    Distance (Mpc)     &    131.8         \\
    Spatial scale(kpc/") &    0.602     \\
    Redshift ({\it z}) &      0.03001     \\
    \noalign{\smallskip}
    \hline
\end{array}
$$
\end{table}

\section{Data}
We used $g$ and $r$ band images of the DECam Legacy Survey (DECaLS) data release 10 (DR10) described in \citet{2019AJ....157..168D}. The DECaLS surveys cover around $\sim$19,000 deg$^2$ area of the sky at high galactic latitudes, mostly in the North and South galactic cap regions. The DECam (Dark Energy Camera) on the Blanco 4m telescope provides the photometry in the $g$, $r$ and $z$ bands with sensitivity from 400$-$1000 nm, and reaches an approximate 5$\sigma$ depth of 23.72, 23.27, and 22.22  AB mag in the g, r, and z bands respectively, for fiducial galaxy with an exponential disk profile having half-light radius of 0.45 arcsec. \citep{2019AJ....157..168D}. It covers the northern galactic cap and southern galactic cap regions with declination $\le$ 32\textdegree and 34\textdegree, respectively. The DR10 comprises coadded images in the $g$, $r$, $i$, and $z$ bands with a pixel scale of 0.262$\arcsec$/pixel. 
  \begin{figure*}[t]
      \centering
      \includegraphics[width=\hsize]{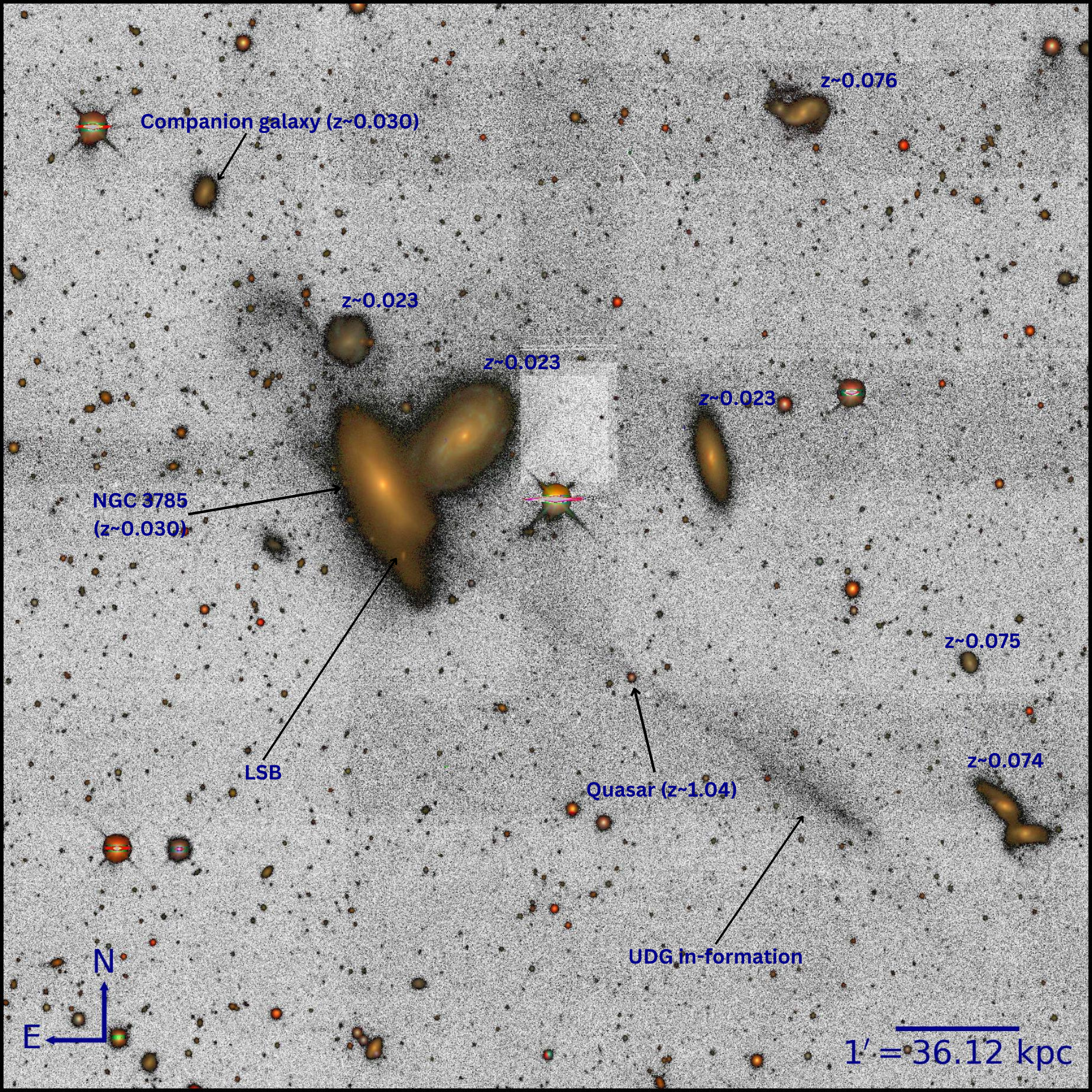}
      \caption{The enhanced tail features of NGC\,3785 are shown in a reversed grey scale wherein the background is light grey, and the emission is grey. The regions of high SNR are shown in color as a $gri$ composite to highlight the different features. The size of the image is $9.1\arcmin \times 9.1\arcmin$ (corresponds to projected field of view of $\sim$ 328.7 $\times$ 328.7 kpc). In this figure, We have marked other prominent galaxies and their corresponding redshift from the SDSS, along with other features such as an LSB (Low surface brightness) galaxy which is described in the discussion section.}
         \label{fig:blendedimage}
  \end{figure*}
\section{Analysis and Results}
The extended tail of NGC\,3785  was discovered serendipitously by one of the co-authors (OB) in the DECaLS imaging. A quick look at PanSTARRS and SDSS images of this galaxy, showed that the tail was faintly visible in these images too. This was enough to convince us that the feature was real and not an image artifact.  To study the extended and faint tail of NGC\,3785, we employed the \texttt{Gnuastro} (GNU Astronomy Utilities) software \citep{2018ascl.soft01009A}. Gnuastro offers a suite of command-line programs designed for efficient data analysis. Among these programs are {\it astnoisechisel, astscript, astcrop, astfits, asttable, astmkcatalog, astsegment,} and others, each serving specific functions, from detecting subtle signals to generating segmentation maps.

\begin{figure*}[t]
     \centering
     \begin{subfigure}[b]{0.5\textwidth}
         \centering
         \includegraphics[width=\textwidth]{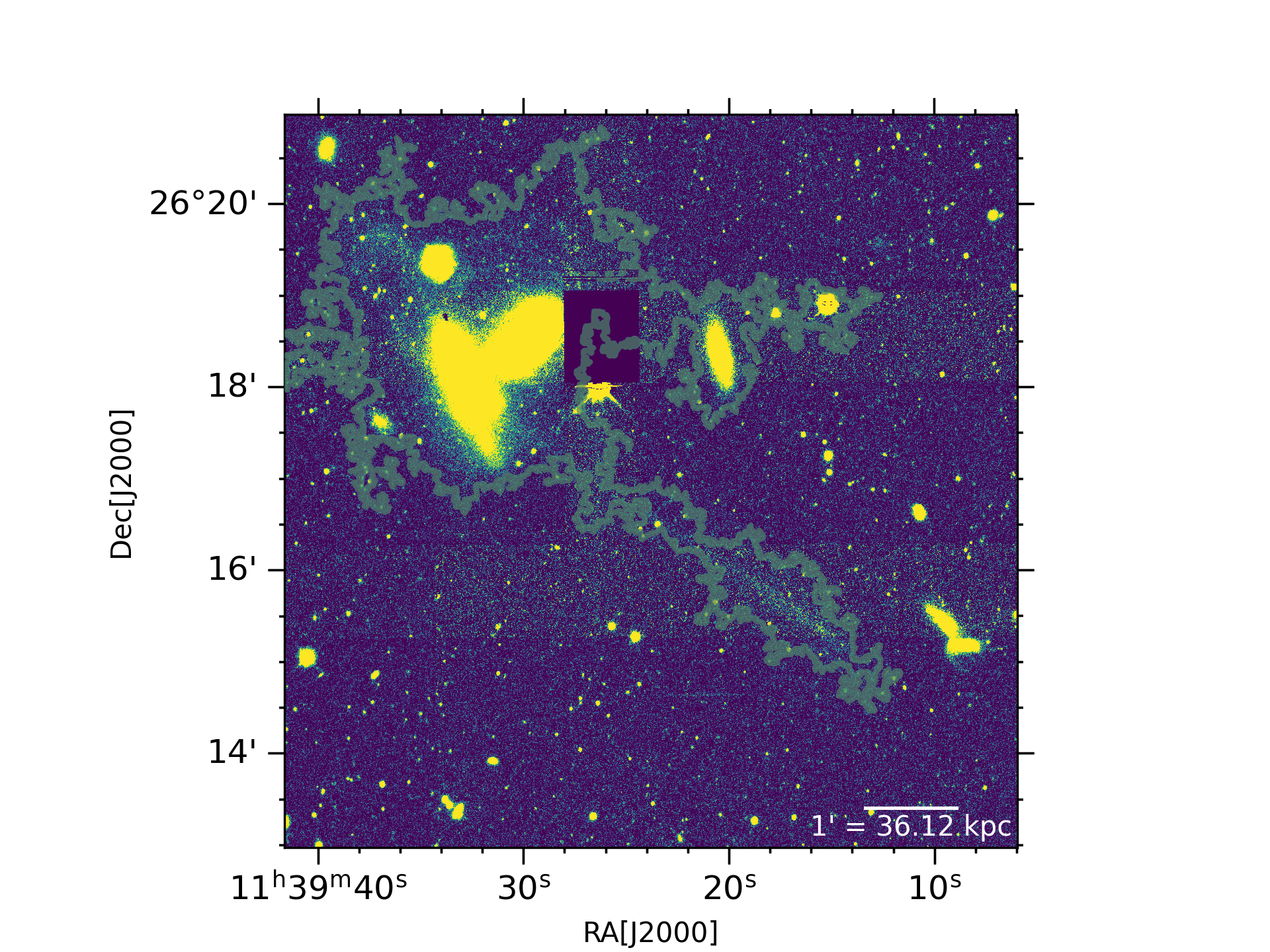}
     \end{subfigure}
     \begin{subfigure}[b]{0.49\textwidth}
         \centering
         \includegraphics[width=\textwidth]{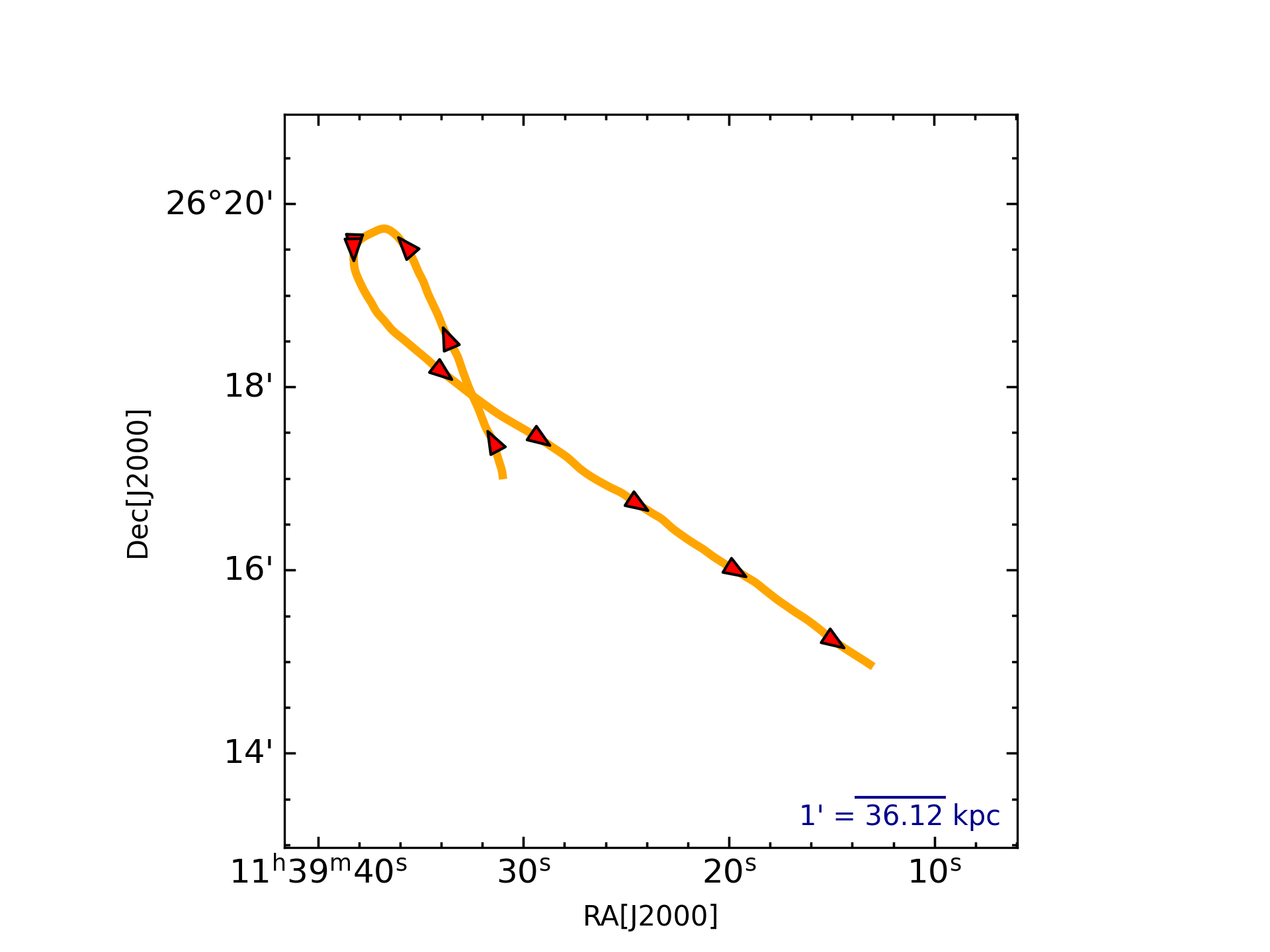}
     \end{subfigure}
        \caption{Left panel: Green contour shows the edge map of faint tail detection on the optical $g$-band image of the galaxy NGC\,3785. The detected edge map serves to define the boundary to the endpoints of the tail.  Right panel: Cubic spline fitting on the data points associated with the straight and looping part of the tail used to measure its length by referencing the edge map and red arrows show the expected trajectory of the satellite galaxy.}
        \label{fig:tail map}
\end{figure*}

The image shown in Figure \ref{fig:blendedimage} was prepared using the {\it astscript-color-faint-grey} script \citep{2024arXiv240103814I}  and shows the larger field of view ($9.1\arcmin \times 9.1\arcmin$) of NGC\,3785 with an extended tail with both straight and curved regions and a companion galaxy {\it SDSS J113939.59+262036.6} at a similar redshift. The {\it astscript-color-faint-grey} script takes three input images representing the RGB channel and combines them after applying a series of non-linear transformations. Then it normalizes the images to create a color-composite image. This offers several parameters such as stretch, colorval, grayval, contrast, weight, etc. to customize the RGB image and enhance the faint features. It uses a default faint grey background to enhance the low surface brightness features. We used the DECaLS DR10 $g$,$r$, and $i$ band images to create the RGB image for better visualization of the extended tail. 

We utilised {\it astnoisechisel} \citep{2019arXiv190911230A} to uncover faint signals within the dataset. This program employs a quantile thresholding method to identify extremely faint and irregular signals embedded within the noise. Subsequently, we employed {\it astsegment} to delineate regions of interest for detailed analysis. {\it astsegment} generates a segmentation map highlighting signals surpassing a specified threshold. It furnishes essential parameters including position (in pixels as well as equatorial coordinates), signal-to-noise ratio, radius (in pixels), magnitude (in nanomaggy), semi-major and semi-minor axes (in pixels), position angle (in degrees), as well as additional parameters like surface brightness (in mag/arcsec$^2$) etc.

To effectively detect the tail, we masked the galaxies surrounding NGC\,3785 to prevent contamination of the light from bright sources. We masked these galaxies with background counts of the field in the $g$-band image because it has more depth to detect faint features as compared to the other bands. Using a masked image alongside a 0.7 detection growth quantile in the $g$-band image, we have successfully identified the faint tail signal and generated an edge map of these detections as shown in Figure \ref{fig:tail map}. This edge map indicates the boundaries of the tail, and we utilized it as a guide to refine the tail's boundaries. Based on visual examination, we marked several points along the tail. Employing the cubic spline fitting technique on these points, we accurately determined the tail's length by referencing the starting and ending points indicated in the galaxy's edge map. The measured projected length encompasses the tail's linear and loop (or curved) segments present in the southwest and northeast of NGC\,3785, respectively, totalling approximately 390 kpc, as illustrated in Figure \ref{fig:tail map}. The northeast and southwest segments are part of only one tidal tail as they follow the same trajectory without any breaks, suggesting that there was only one tidal event. It seems that the satellite galaxy has undergone a tidal interaction with the NGC 3785, starting from the northeast, making a loop, and then going towards the southwest by following the same trajectory. Therefore, this indicates that a single tidal event created the tidal tail, and this path is clearly visible in Figure \ref{fig:blendedimage}. This observation makes the tail of NGC\,3785 the longest tidal tail discovered to date.  
 
Once the tail's borders are defined, the subsequent task involves identifying regions or clumps within the tail region. The purpose of this exercise is to measure the color and surface brightness of these clumps, for further analysis. We employed the {\it astsegment} task within Gnuastro to identify clumps and generate a segmentation map from the detections produced by the {\it astnoisechisel} task. Using these clump detections, we compiled a catalogue containing their IDs, RA, Dec, magnitude, surface brightness, and signal-to-noise ratio. In total, 903 clumps were detected in the entire $9.1\arcmin \times 9.1\arcmin$ $g$-band image, encompassing both the tail and other regions. However, our focus is on isolating the clumps within the tail, a process achieved through manual visual inspection, resulting in a reduced count of 100 clumps.

\begin{figure}
  \centering
  \includegraphics[width=0.5\textwidth]{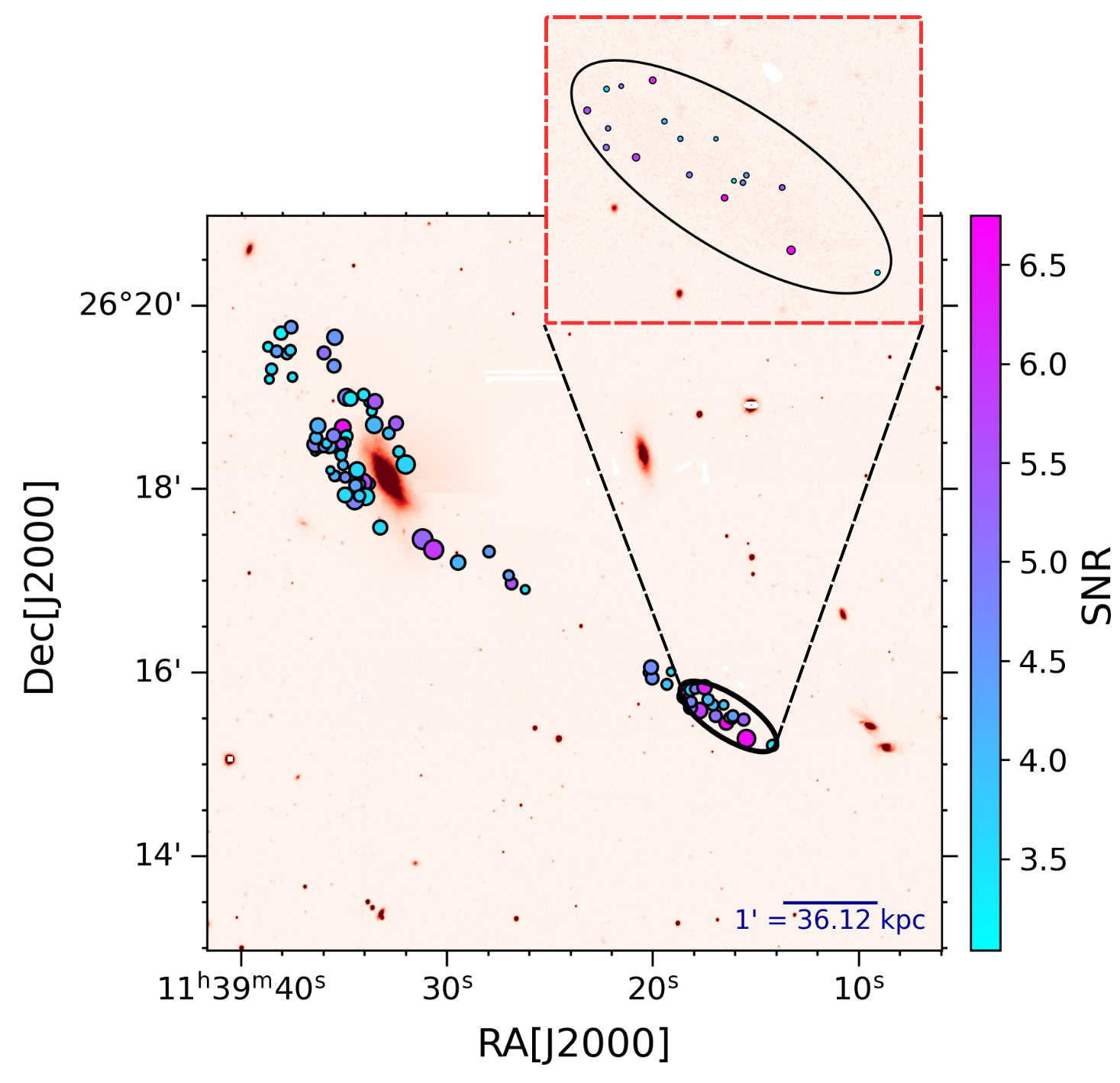}
  \caption{This figure shows the clumps we detect in the tail and the inset shows the zoomed-in view of the concentrated clumps at the tip of the tail. Each clump is colored by its signal-to-noise ratio in the $g$-band. We have taken the elliptical aperture around the eighteen clumps to study their photometric properties such as $g-r$ color, absolute magnitude, and stellar mass. Here, the size of the circle represents the area of the clump in pixels.}
  \label{fig:variation}
\end{figure}

Among the 100 clumps detected in the tail, we are left with 84 clumps after applying a signal-to-noise threshold of 3 as shown in Figure \ref{fig:variation}. These 84 clumps are more prominent in the $g$-band image compared to the $r$-band image. When we superimposed the $g$-band detected clumps onto the $r$-band image, we observed that 48 of them had detectable $r$-band emission, while the other 36 do not have any emission in the $r$-band. These clumps are associated with the faint part of the tail. Although they have a good signal-to-noise ratio, their full width at half maximum (FWHM) is smaller than the point spread function (PSF) in the $g$-band (1.29\arcsec), resulting in an unrealistic $g-r$ color. Therefore, we cannot use these clumps for the photometric analysis such as the $g-r$ color map. Moreover, Gnuastro provides a surface brightness of these clumps, which covers a range from approximately 23.92 $\pm$ 0.45 to 26.55 $\pm$ 0.29 mag$/$arcsec$^2$ in the $g$-band. These numbers can be compared to those from a recent study by \cite{2024arXiv240720483M},  who identified 63 tidal streams in nearby galaxies within 100 Mpc using DECaLS $g$,$r$, and $z$ band data. These tidal streams have an average surface brightness range of 24.86 $<$ $\mu_{g}$ $<$ 28.05 mag$/$arcsec$^2$ with 1$\sigma$ uncertainty. The average projected length of these tidal streams is approximately 50 kpc, which is small compared to the extended tail we detected.

\section{Discussion}
Our analysis reveals the longest tail ever observed within this isolated environment which has a projected length of 390 kpc and the deprojected length could be even larger. It is crucial to note that this tail is not uniformly straight; instead, it comprises both a linear segment and a looping section. Upon visual inspection of the image, two potential scenarios emerge. First, we anticipate the tail initiating from the lower right corner of the image, progressing directly toward the galaxy, forming a loop, and concluding just below the galaxy. In an alternative scenario, a companion low-mass galaxy appears to orbit around the central high-mass galaxy before proceeding along a linear trajectory towards the southwest, as shown in the right panel of Figure \ref{fig:tail map}. The latter scenario seems more likely for this particular case, an idea we will explore further in this section.

While detecting clumps on the tail, we have noticed a clump at the southeast end of the tail marked as LSB in Figure \ref{fig:blendedimage}. It might be the companion galaxy that interacted with NGC\,3785 and ended up as a low surface brightness galaxy, which is feasible according to the first scenario. In order to calculate the stellar mass of this clump, We used the relation given by \cite{2020AJ....159..138D}, which provides a relation between the stellar $M/L$ ratio and the optical colors for low-surface brightness galaxies. Using this relation, we obtained the stellar mass of the clump as $10^{8.42} M_{\odot}$in the $g$-band having color $g-r = 0.90$. This has a surface brightness of 23.843 $\pm$ 0.071 mag arcsec$^{-2}$ in $g$-band.  

\cite{1992Natur.360..715B}, \cite{2010A&A...521A..20G} and \cite{2023ApJ...954L...2C} are the studies that also indicate that after an interaction, the tidally stripped galaxy forms small dwarf galaxies which favours the second scenario. The simulations done by \cite{1992Natur.360..715B} suggest that dwarf galaxies form within the tidal tails of the galaxies that have undergone interactions. These structures can form in the outer regions of the tidal tail as a result of either the Jeans instabilities within the gas or due to a significant fraction of the stellar material being ejected from the progenitor disk to the tidal tail \citep{2012MNRAS.419...70K}. Depending upon the surface brightness, these can result in low surface brightness galaxies (LSB), ultra-diffuse galaxies (UDGs) or tidal dwarf galaxies (TDGs). UDGs have a very low central surface brightness ($\mu$(g,0) > 24 mag arcsec$^{-2}$), but their half-light radius is comparable to typical galaxies.

In Figure \ref{fig:blendedimage}, one can see that the bottom end of the tail (southwest) has more concentration of clumps, and these are also visible in Figure \ref{fig:variation} as well, where most of the clumps are concentrated at the end of the southwest part of the galaxy. If we consider the second scenario and the simulations mentioned above, it might be possible that some structure has started forming at the tip of the tail. We used the Two Point Correlation Function (TPCF) to determine if there is actual clustering and to uncover the distribution of these concentrated clumps. TPCF measures the clumpiness of the data points \citep{1980lssu.book.....P} and is calculated as the ratio of the data pairs to the randomly distributed pairs at a given distance in the same bin. As the angular separation between two data points increases, TPCF decreases, resulting in a distortion in the hierarchical distribution \citep{2021MNRAS.507.5542M}. It implies that higher TPCF values correspond to a higher probability of clustering. For the calculation of angular TPCF in this field, we employed the Landy-Szalay (LS) estimator \citep{1993ApJ...412...64L}. We observed that the majority of the clumps having high TPCF (>1) are located near the galaxy NGC\,3785  and around the tip of its tail (southwest part). It points towards a clustering of clumps in this region.

\begin{figure}
    \hspace{-0.5cm}
    \centering
    \includegraphics[width=0.5\textwidth]{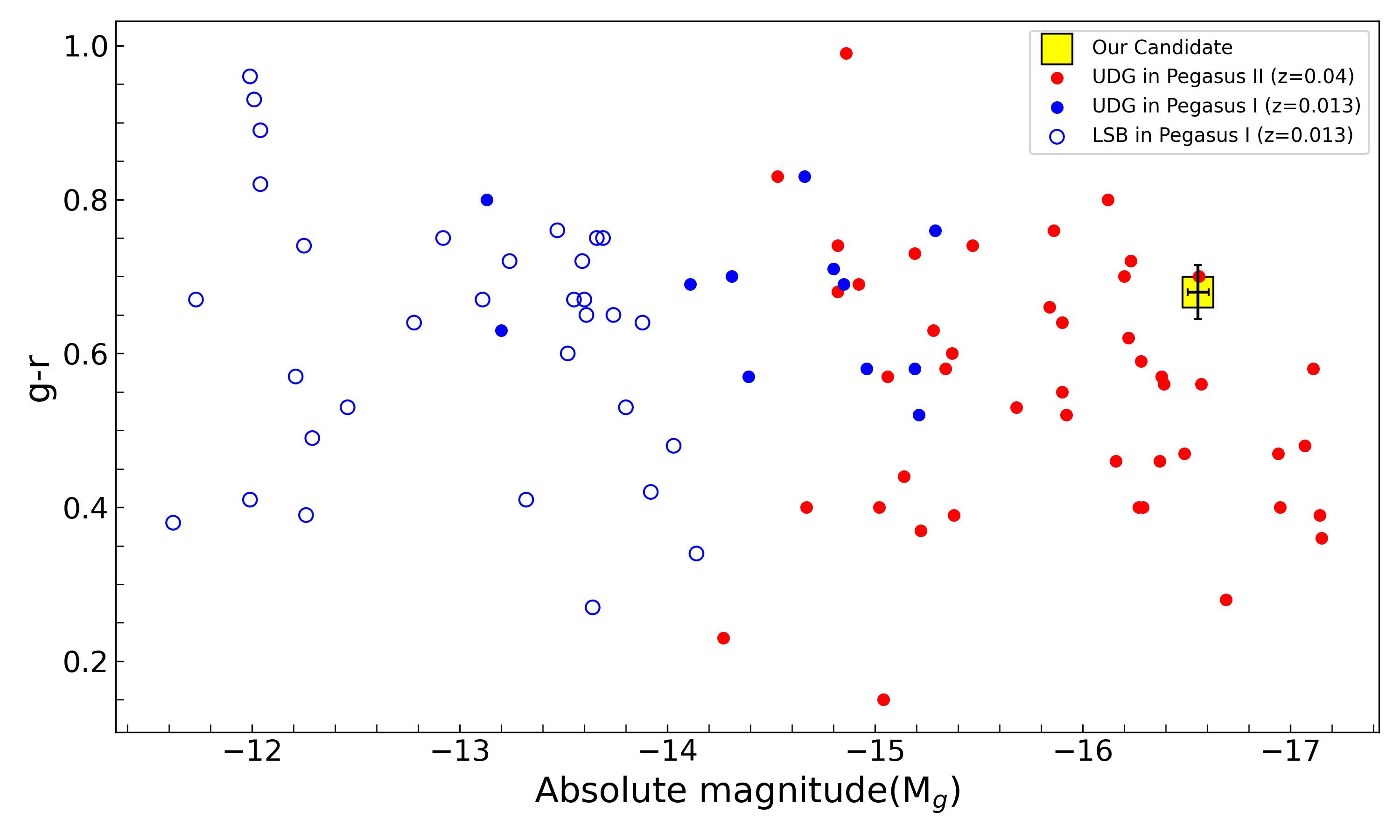}
    \caption{We have recreated the $g-r$ vs absolute magnitude plot from \cite{2017ApJ...846...26S} to determine whether the faint galaxy at the tip of the tail is LSB or a UDG. The data points of LSB and UDG in Pegasus I and II are taken from \cite{2017ApJ...846...26S}. Our candidate (faint galaxy), is depicted as a yellow square box with error bars, which lies in the region populated by UDGs.}
    \label{fig:shi}
\end{figure}

To further investigate the nature of these clustered clumps, we took the elliptical aperture around those clumps as shown in Figure \ref{fig:variation}, and performed the elliptical aperture photometry using $\it{Photutils}$ v1.9.0 \citep{2023zndo...8248020B} to calculate the magnitude and surface brightness in the background subtracted $g$ and $r$ band images. From photometric analysis, we found that $g-r$ color corresponds to a 0.68 $\pm$ 0.05 mag with an absolute magnitude (M$_g$) of -16.55 $\pm$ 0.03 mag and surface brightness of 27.04 $\pm$ 0.03 mag arcsec$^{-2}$. Using the relation between the $M/L$ ratio and the optical colors for low-surface brightness galaxies \citep{2020AJ....159..138D}, the $g-r$ color of the elliptical aperture corresponds to a stellar mass of $10^{8.82} M_{\odot}$. This aperture has very low surface brightness, which satisfies the criteria of UDGs. To determine whether this aperture is associated with the formation of a UDG or LSB galaxy, we replicate Figure 8 from \cite{2017ApJ...846...26S}. This figure plots the relationship between $g-r$ color and absolute magnitude, allowing for differentiation between LSBs, UDGs, and typical galaxies. We placed our candidate on the same plot and observed that it is located near the ultra-diffuse galaxies (UDGs) of Pegasus II (at $z =0.04$) as shown in Figure \ref{fig:shi}. However, our candidate  UDG is located within the red sequence, but it is positioned near the boundary separating the red and blue sequences, and this boundary is typically for normal SDSS galaxies. The UDGs from  \cite{2017ApJ...846...26S} are already formed and normally isolated, i.e. they do not show any signature of ongoing interactions. In our case, the candidate  UDG is still associated with a tidal tail as the gas-stripping mechanisms such as tidal stripping play a crucial role here, causing most of the gas to be stripped away from the satellite galaxy (\cite{2004A&A...425..813B}, \cite{1995AJ....110..140H}, \cite{1972ApJ...178..623T}) as it traveled several kiloparsecs to the UDG position. As a result, it lacks sufficient gas, placing it close to a division of red and blue sequence boundary. It seems that tidal stripping for the candidate UDG is still ongoing, suggesting that it is in the process of becoming a UDG. Due to its diffuse nature and low surface brightness, measuring the effective radius for this UDG is non-trivial. Several studies (For e.g. \cite{10.1093/pasj/psx051,2018ApJ...866L..11B,2023MNRAS.518.2497Z}) highlights the association of tidal features with UDGs. This supports the idea that tidal interactions play a vital role in the formation and evolution of UDGs.

The clumps detected on the tidal tail are more prominent in the $g$-band, suggesting that they are blue in color. These clumps are exclusively related to the tail part, as we have discarded any clumps that are part of other bright sources or galaxies while analyzing the tail. Also, NGC\,3785 is situated in an isolated environment with no other galaxy present at the same redshift in its field. While S0 galaxies typically have lower star formation due to their lack of gas, the presence of a long tail associated with this S0 galaxy indicates that it may have undergone interaction with a gas-rich galaxy in the past, leaving behind a trail of star formation.

\section{Summary}
We have discovered an exceptionally long and faint tail in an isolated environment, likely originating from galaxy interactions, particularly associated with NGC\,3785. This tail is seen prominently in the $g$-band image of the galaxy. Employing an edge map, we measured its length, utilizing the cubic spline fitting method to accommodate its mix of straight and curved segments. Our analysis reveals that the tail has a projected length of approximately 390 kpc, the longest tail observed in an isolated environment. It has a similar length to the previously discovered longest tail by \cite{2023MNRAS.524.1431Z}, who noted a tail with a projected length of 380 kpc resulting from the tidal interaction between the Kite and the Mrk 0926 galaxies.

We utilized the tail segment identified in our analysis to locate star-forming clumps, and after filtering out clumps unrelated to the tail, we were left with 84 clumps in the tail with a signal-to-noise ratio above three in the $g$  band. These 84 clumps revealed that the clumps in the tail region are blue and thus prominent in the $g$-band, indicative of a possible interaction between NGC\,3785 and a gas-rich galaxy, resulting in a trail of stars. Moreover, our examination suggests the imminent emergence of an Ultra Diffuse Galaxy (UDG) at the tail's terminus.

After thoroughly examining the images, we have detected a quasar, {\it SDSS J113923.47+261630.5}  ($z\sim 1.04$),  situated behind the galaxy's tail marked in Figure \ref{fig:blendedimage}. From the SDSS spectra for this quasar \citep{2023ApJS..267...44A}, we investigated if it could assist us in identifying any absorption features caused by the tidal stream. Regrettably, the spectra are quite noisy, posing challenges in distinguishing faint absorption lines from background noise. In the future, deep spectroscopic observations of the quasar and IFU observations of the bottom of the tail can help to provide more insights into the formation of the UDG at the tip of the tail. Upcoming space and ground-based deep surveys using the recently launched Euclid Space Telescope (Euclid), and the Legacy Survey of Space and Time (LSST) to soon commence at the Rubin Observatory will provide more insights into the fascinating world of tidal tails and low surface brightness galaxies.

\begin{acknowledgements}
We thank the anonymous referee for the thoughtful review, which improved the impact and clarity of this work. YW acknowledges support from the Department
of Atomic Energy, Government of India, under project
12-RD-TFR-5.02-0700. The photometry analysis in this work was partly done using GNU Astronomy Utilities (Gnuastro,
ascl.net/1801.009) version 0.22. Work on Gnuastro has been
funded by the Japanese MEXT scholarship and its Grant-in-Aid for Scientific
Research (21244012, 24253003), the European Research Council (ERC) advanced grant 339659-MUSICOS, and from the Spanish Ministry of Economy
and Competitiveness (MINECO) under grant number AYA2016-76219-P. M.A
acknowledges the financial support from the Spanish Ministry of Science and
Innovation and the European Union - NextGenerationEU through the Recovery and Resilience Facility project ICTS-MRR-2021-03-CEFCA and the grant
PID2021-124918NA-C43.  The Legacy Surveys consist of three individual and complementary projects: the Dark Energy Camera Legacy Survey (DECaLS; Proposal ID 2014B-0404; PIs: David Schlegel and Arjun Dey), the Beijing-Arizona Sky Survey (BASS; NOAO Prop. ID 2015A-0801; PIs: Zhou Xu and Xiaohui Fan), and the Mayall z-band Legacy Survey (MzLS; Prop. ID 2016A-0453; PI: Arjun Dey). DECaLS, BASS and MzLS together include data obtained, respectively, at the Blanco telescope, Cerro Tololo Inter-American Observatory, NSF’s NOIRLab; the Bok telescope, Steward Observatory, University of Arizona; and the Mayall telescope, Kitt Peak National Observatory, NOIRLab. Pipeline processing and analyses of the data were supported by NOIRLab and the Lawrence Berkeley National Laboratory (LBNL). The Legacy Surveys project is honored to be permitted to conduct astronomical research on Iolkam Du’ag (Kitt Peak), a mountain with particular significance to the Tohono O’odham Nation. 
\end{acknowledgements}

\bibliography{ngc3785}

\end{document}